\documentclass[aps,prl,showpacs,superscriptaddress,twocolumn,longbibliography]{revtex4-2}
\pdfoutput=1

\usepackage{amsfonts}
\usepackage{subfigure}
\usepackage{amsmath}
\usepackage{amssymb}
\usepackage{amsbsy} 
\usepackage{graphicx}

\def\be{\begin{equation}} \def\ee{\end{equation}}
\def\bea{\begin{eqnarray}} \def\eea{\end{eqnarray}}
\def\nn{\nonumber}

\def\rw{\rightarrow}
\def\bk{{\bf k}}

\def\ra{\rangle}
\def\la{\langle}

\begin{document}

\title{Non-Bloch PT symmetry breaking: Universal threshold and dimensional surprise}

\author{Fei Song}
 \affiliation{ Institute for
Advanced Study, Tsinghua University, Beijing,  100084, China }

\author{Hong-Yi Wang}
\affiliation{ Institute for
Advanced Study, Tsinghua University, Beijing,  100084, China }
\affiliation{ School of Physics, Peking University, Beijing, 100871, China  }

\author{Zhong Wang} \altaffiliation{ wangzhongemail@gmail.com }
\affiliation{ Institute for
Advanced Study, Tsinghua University, Beijing,  100084, China }


\begin{abstract}

In the presence of non-Hermitian skin effect, non-Hermitian lattices generally have complex-valued eigenenergies under periodic boundary condition, but they can have \emph{non-Bloch PT symmetry} and therefore completely real eigenenergies under open boundary condition. This novel PT symmetry and its breaking have been experimentally observed in one dimension. Here, we find that non-Bloch PT symmetry in two and higher dimensions exhibits drastically different behaviors compared to its one-dimensional counterpart. Whereas Bloch PT breaking and one-dimensional non-Bloch PT breaking generally have nonzero thresholds in the large-size limit, the threshold of two and higher-dimensional non-Bloch PT breaking universally approaches zero as the system size increases. A \emph{product measure}, namely the product of bare non-Hermiticity and system size, is introduced to quantify the PT breaking tendency. This product being small is required for the perturbation theory to be valid, thus its growth with system size causes the breakdown of perturbation theory, which underlies the universal threshold. That the universal behaviors emerge only in two and higher dimensions indicates an unexpected
interplay among PT symmetry, non-Hermitian skin effect, and spatial dimensionality. Our predictions can be confirmed on experimentally accessible platforms.

\end{abstract}
\maketitle

\emph{Introduction.--}In the standard quantum mechanics of closed systems, the Hamiltonians are always Hermitian. The time evolution of open systems is, however, often generated by effective non-Hermitian Hamiltonians \cite{Ashida2020,Bergholtz2020RMP,rotter2009non}. While complex-valued eigenenergies are natural consequences of gain and  loss, a prominent class of non-Hermitian Hamiltonians can have purely real eigenenergies when the non-Hermiticity is below a threshold \cite{bender1998real,bender2007making}. They are known as parity-time (PT) symmetric Hamiltonians, and the real-to-complex transition is called PT symmetry breaking \cite{Ozdemir2019review,el2018non,Miri2019review}. PT symmetry and its breaking have many intriguing consequences, e.g., single-mode lasing \cite{feng2014singlemode,hodaei2014PT}, nonreciprocal transmission \cite{feng2011nonreciprocal,ruter2010observation,peng2014parity}, and unidirectional invisibility \cite{lin2011unidirectional,regensburger2012parity}.

Independent of the PT symmetry, non-Hermitian topology has recently been attracting growing attention. Remarkably, the bulk-boundary correspondence is drastically modified by the non-Hermitian skin effect (NHSE) \cite{yao2018edge,yao2018chern,kunst2018biorthogonal,lee2018anatomy,alvarez2017}, namely that all the energy eigenstates are exponentially squeezed to the boundary. Owing to this failure of Bloch band picture, the edge states correspond to the non-Bloch topological invariants defined in the generalized Brillouin zone (GBZ) \cite{yao2018edge,yao2018chern,Yokomizo2019,Yang2019Auxiliary,Deng2019,Kawabata2020nonBloch, Longhi2019nonBloch,Longhi2019Probing,
Song2019,Song2019real,Yi2020,Li2020critical,Longhi2020chiral,liu2019second,Lee2020Unraveling,Liu2020helical, Yokomizo2020BdG,Yang2020BdG,Yokomizo2020review}, which underlies the non-Bloch band theory \cite{yao2018edge,Yokomizo2019,Kawabata2020nonBloch,Yokomizo2020review}. Recent experiments have confirmed the novel bulk-boundary correspondence of non-Hermitian systems \cite{Helbig2019NHSE,Xiao2019NHSE,Weidemann2020,Hofmann2020,Ghatak2019NHSE,Palacios2020}.

Very recently, an intriguing interplay has been found between PT symmetry and NHSE in one-dimensional (1D) systems \cite{Longhi2019nonBloch,Longhi2019Probing}. In the presence of NHSE, the non-Bloch bands consisting of eigenstates under open-boundary condition (OBC) can have PT symmetry, though the Bloch bands under periodic-boundary condition (PBC) cannot. Thus, NHSE becomes a mechanism of PT symmetry in periodic lattices that lies outside the familiar Bloch band framework. This NHSE-induced PT symmetry, dubbed \emph{non-Bloch PT symmetry} \cite{Longhi2019nonBloch,Longhi2019Probing}, has been experimentally observed recently \cite{Xiao2020nonBloch}.

In this paper, we find that non-Bloch PT symmetry in two and higher dimensions has drastically different and unexpected behaviors compared to the 1D cases. It turns out that the threshold of 2D non-Bloch PT symmetry breaking universally approaches zero as the system size increases. Even for an infinitesimal non-Hermiticity, a large proportion of eigenenergies undergo the real-to-complex transitions as the system size increases. This feature is in sharp contrast to the PT transitions of Bloch bands, which generally have nonzero thresholds; even when fine tuned to thresholdless points, an infinitesimal non-Hermiticity can at most cause transitions of an infinitesimal proportion of eigenenergies, regardless of the system size \cite{Barashenkov2013,Konotop2016nonlinear}. Notably, non-Bloch PT breaking in 1D also has a size-independent (generally nonzero) threshold at large size. Thus, our finding reveals an unexpected interplay between non-Bloch bands and spatial dimensionality.

\begin{figure}
\includegraphics[width=4cm, height=4cm]{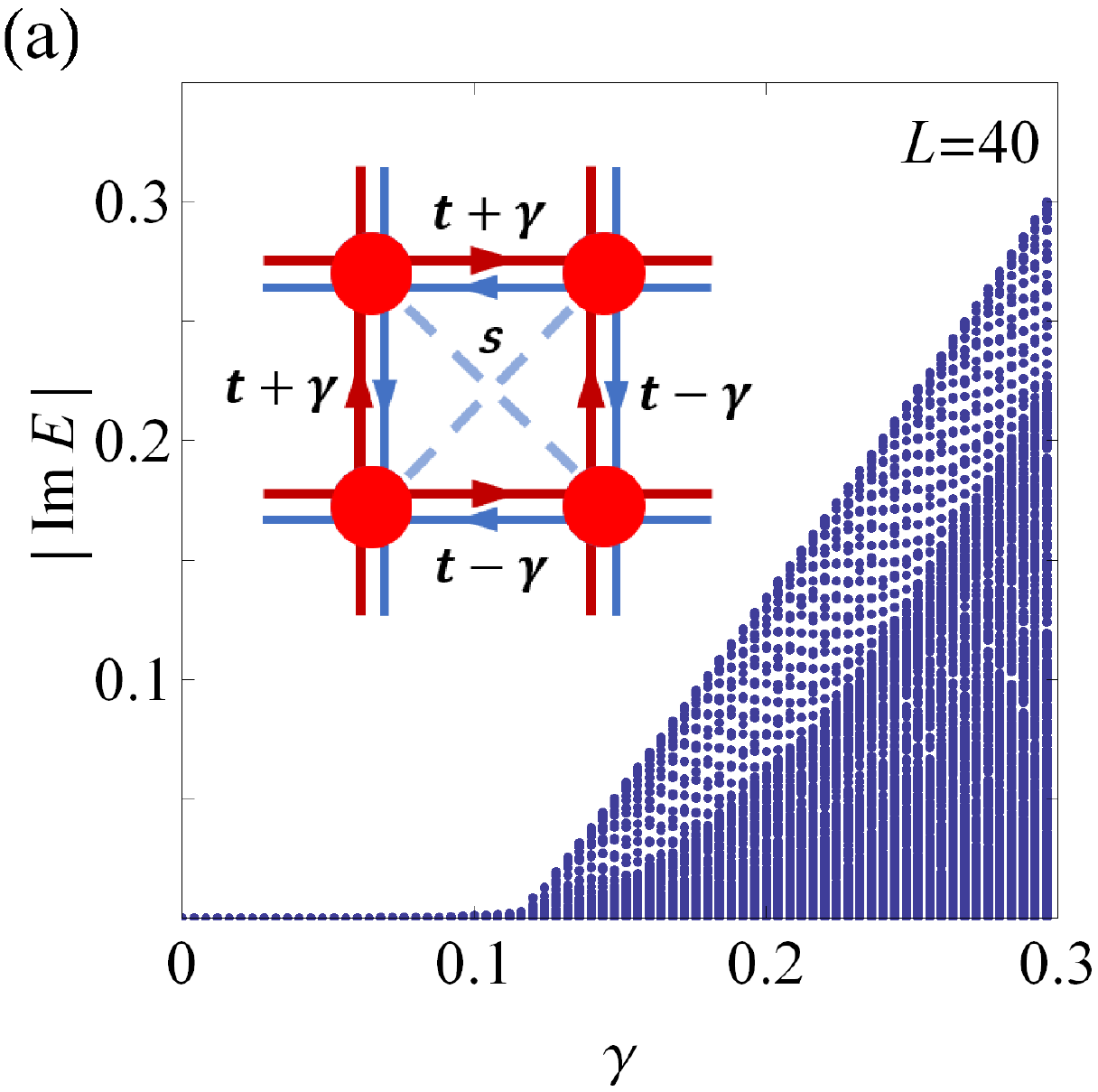}
\includegraphics[width=4cm, height=4cm]{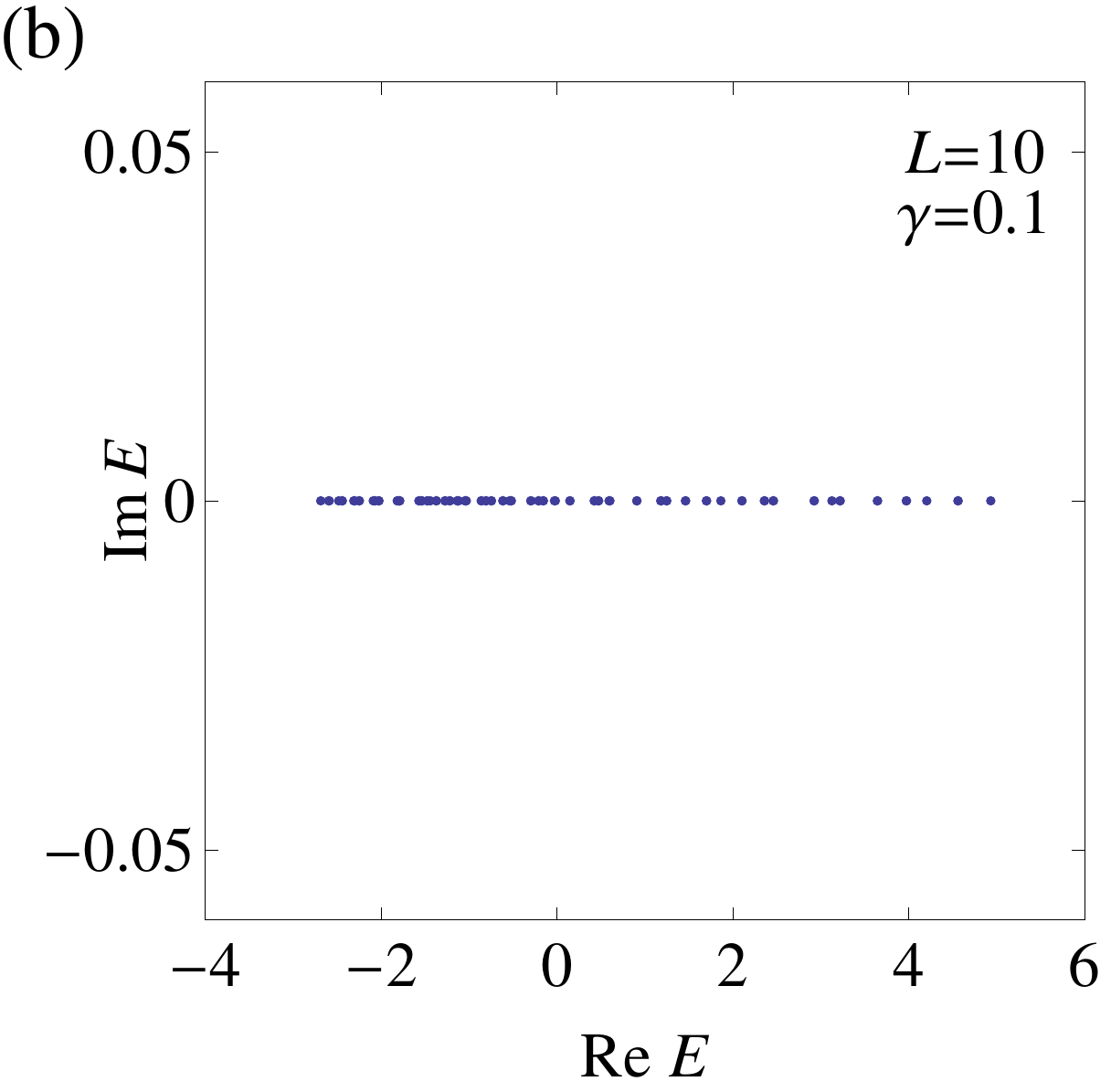}
\includegraphics[width=4cm, height=4cm]{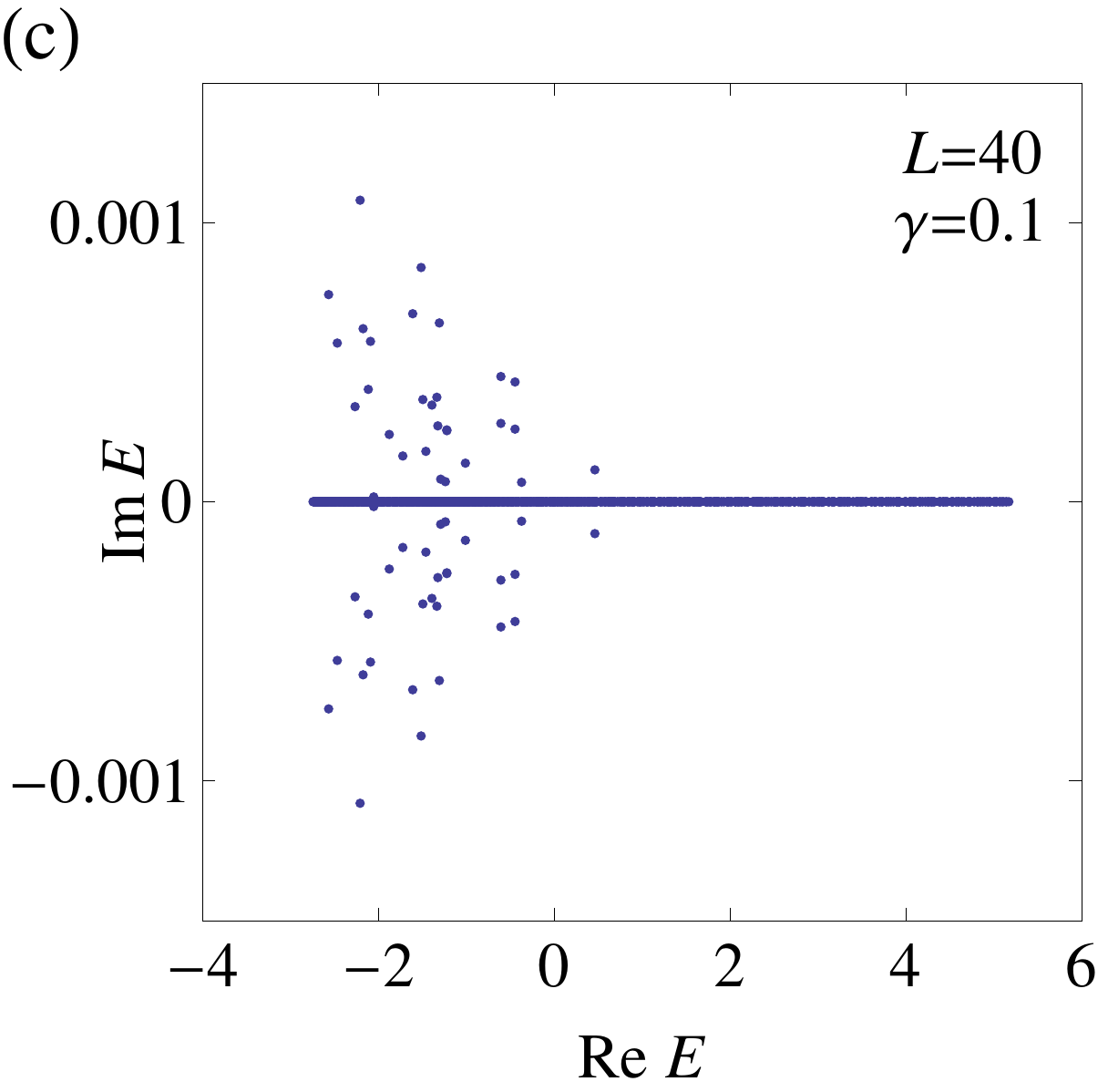}
\includegraphics[width=4cm, height=4cm]{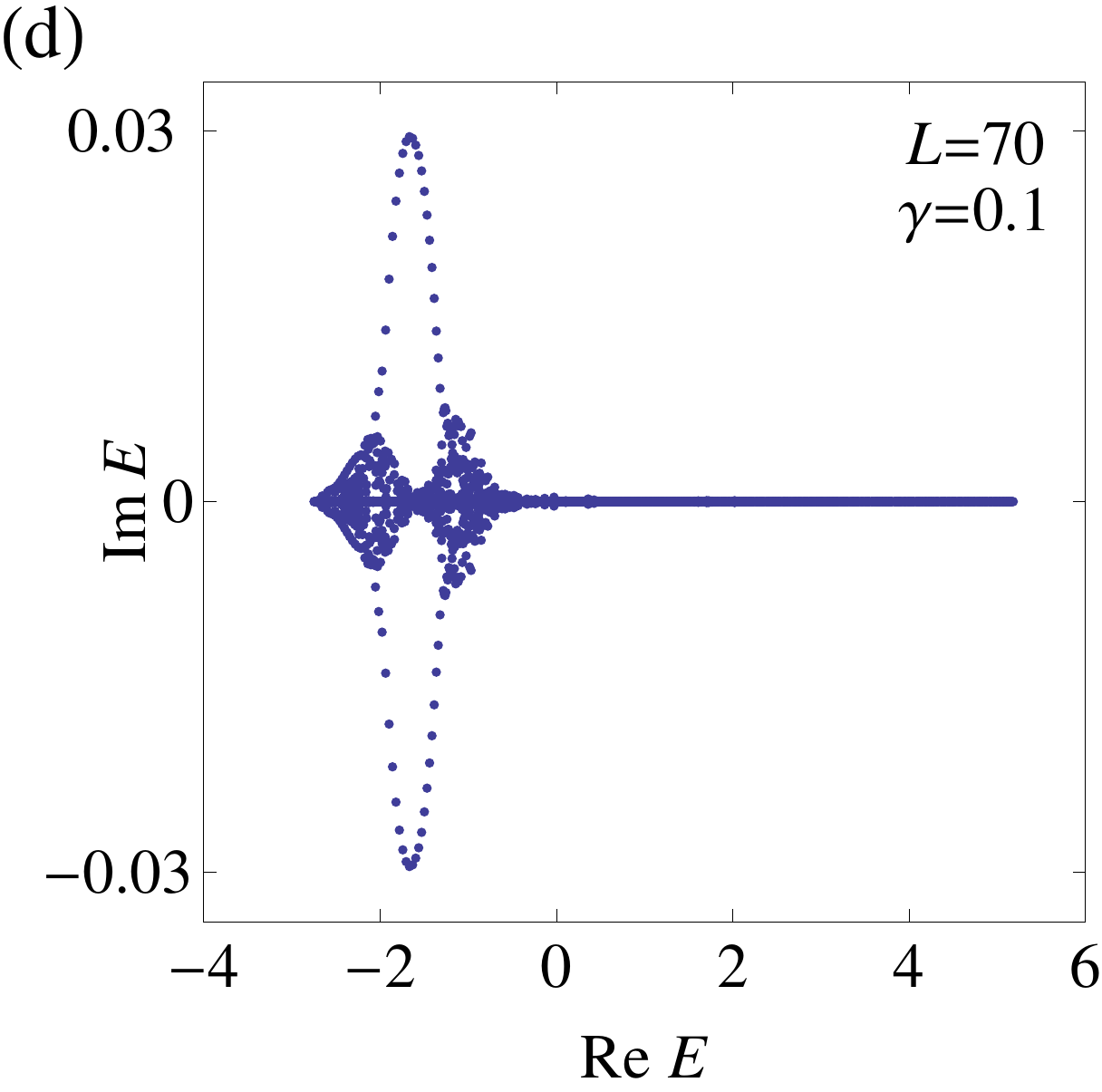}
\caption{Energy spectrums of the single-band model Eq.~(\ref{H}) for $L\times L$ squares. (a) Absolute values of imaginary parts of all eigenenergies for $L=40$ with varying $\gamma$. The inset shows the hoppings in Hamiltonian. (b,c,d) Eigenenergies for $L=10,40,70$, respectively, with fixed $\gamma=0.1$. Other parameters are fixed as $t=1$ and $s=0.3$.  } \label{fig1}
\end{figure}

\emph{Universal threshold.--}We consider a simple 2D non-Hermitian lattice with hoppings shown in Fig.~\ref{fig1}(a), whose Bloch Hamiltonian is
\bea
H(\bk)=(t-\gamma)e^{ik_x}+(t+\gamma)e^{-ik_x}+(t-\gamma)e^{ik_y}+ \nn\\(t+\gamma)e^{-ik_y}+s(e^{ik_x}+e^{-ik_x})(e^{ik_y}+e^{-ik_y}), \label{H}
\eea where $\bk=(k_x,k_y)$ and $t,\gamma,s$ are real parameters. As a single-band model, its PBC eigenenergies are just $H(\bk)$, which are generally complex-valued. The OBC eigenenergies are entirely different because of the NHSE. For the simplest case $s=0$, all the eigenstates are localized at a corner of the system, taking the form of $\psi(x,y)\sim (\beta_x)^x (\beta_y)^y$ with $|\beta_x|=|\beta_y|=\sqrt{\frac{t+\gamma}{t-\gamma}}$. This is readily seen by a similarity transformation akin to that used for the non-Hermitian SSH model\cite{yao2018edge}. An equivalent statement is that the GBZ is the 2D torus $\{(\beta_x,\beta_y): |\beta_x|=|\beta_y|=\sqrt{\frac{t+\gamma}{t-\gamma}} \}$. Although analytical formula is unavailable when $s\neq 0$, the NHSE is still expected.

The real-space Hamiltonian has a generalized PT symmetry $\mathcal{K}H\mathcal{K}=H$, where $\mathcal{K}$ stands for complex conjugation, as the hoppings are all real-valued\cite{Bender2002}. Accordingly, the OBC eigenenergies are real-valued for sufficiently small $\gamma$. As usual, PT breaking transition occurs at some threshold. An example is shown in Fig.~\ref{fig1}(a), where the threshold $\gamma$ is close to $0.1$. This is reminiscent of 1D non-Bloch PT transitions \cite{Longhi2019nonBloch,Longhi2019Probing,Xiao2020nonBloch}.

The surprise comes when we study the size dependence. Taking $\gamma=0.1$ for example, we find unbroken PT phase at $L=10$ [Fig.~\ref{fig1}(b)]. However, at $L=40$ the PT symmetry is broken, and at $L=70$ even more proportion of complex-valued eigenenergies appear [Fig.~\ref{fig1}(c,d)]. Similarly, taking other values of $\gamma$, no matter how small, we always see PT breaking by increasing $L$. This means that the threshold approaches zero as size increases. This behavior drastically differs from Bloch (NHSE-free) PT breaking and non-Bloch PT breaking in 1D \cite{Longhi2019nonBloch,Longhi2019Probing,Xiao2020nonBloch}. In both cases, the threshold generally converges to a nonzero constant as size increases.

\begin{figure}
\includegraphics[width=4cm, height=3.45cm]{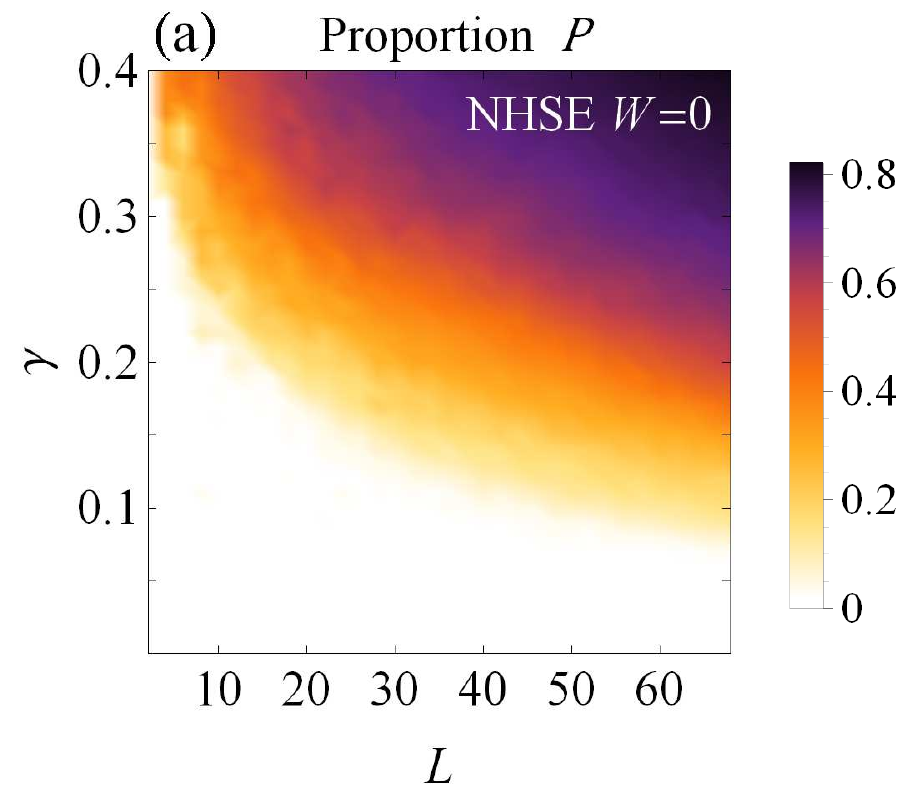}
\includegraphics[width=4cm, height=3.45cm]{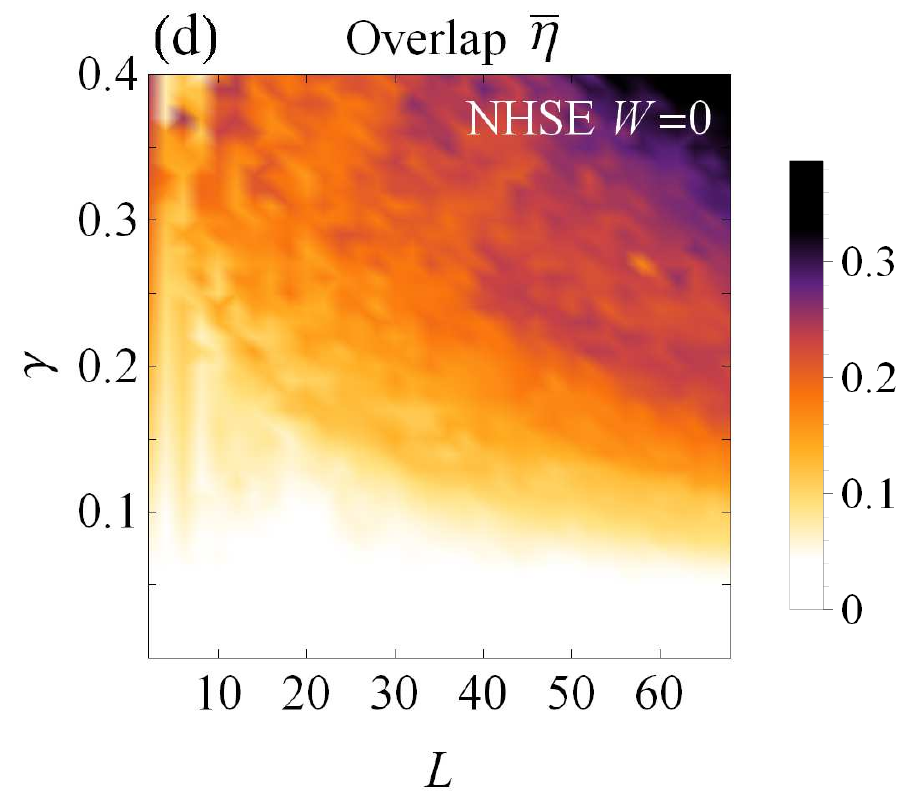}
\includegraphics[width=4cm, height=3.45cm]{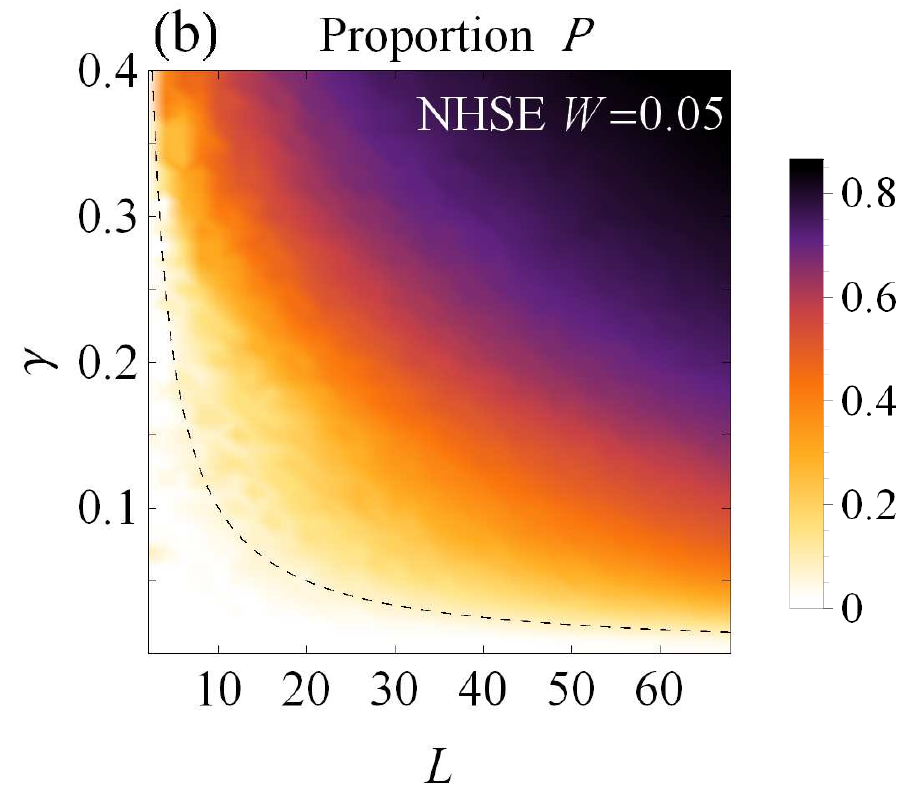}
\includegraphics[width=4cm, height=3.45cm]{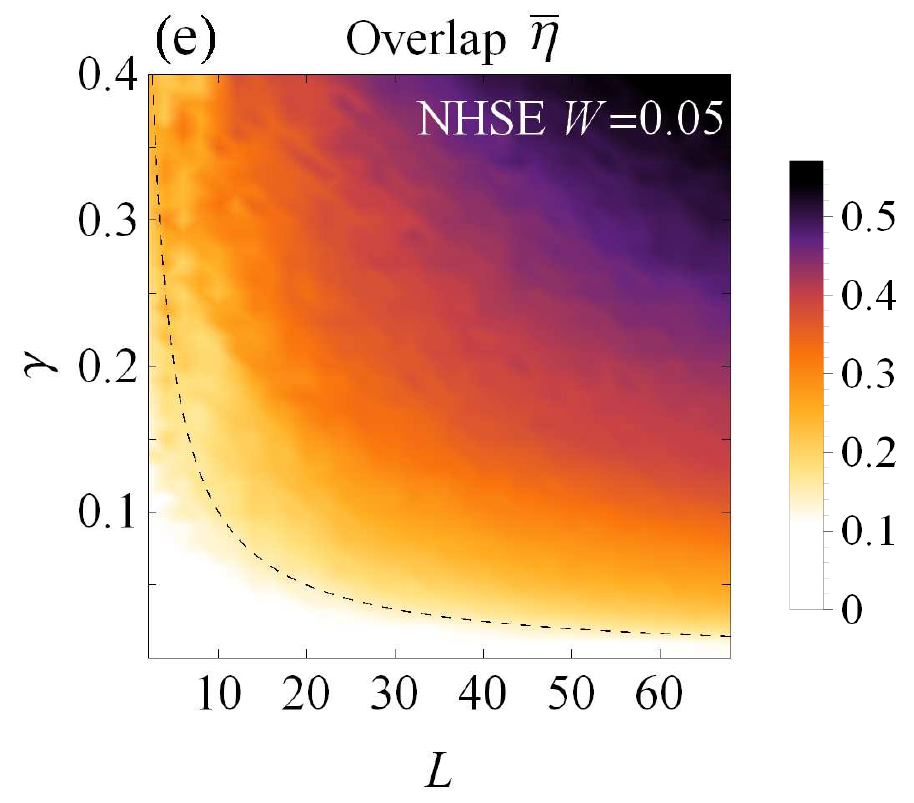}
\includegraphics[width=4cm, height=3.45cm]{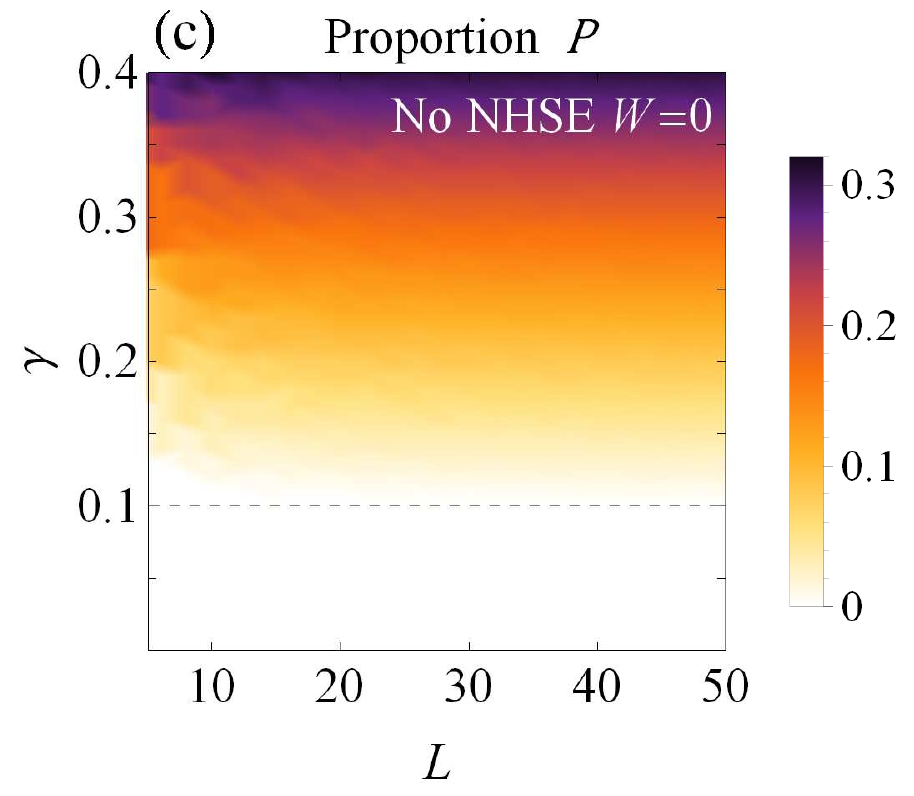}
\includegraphics[width=4cm, height=3.45cm]{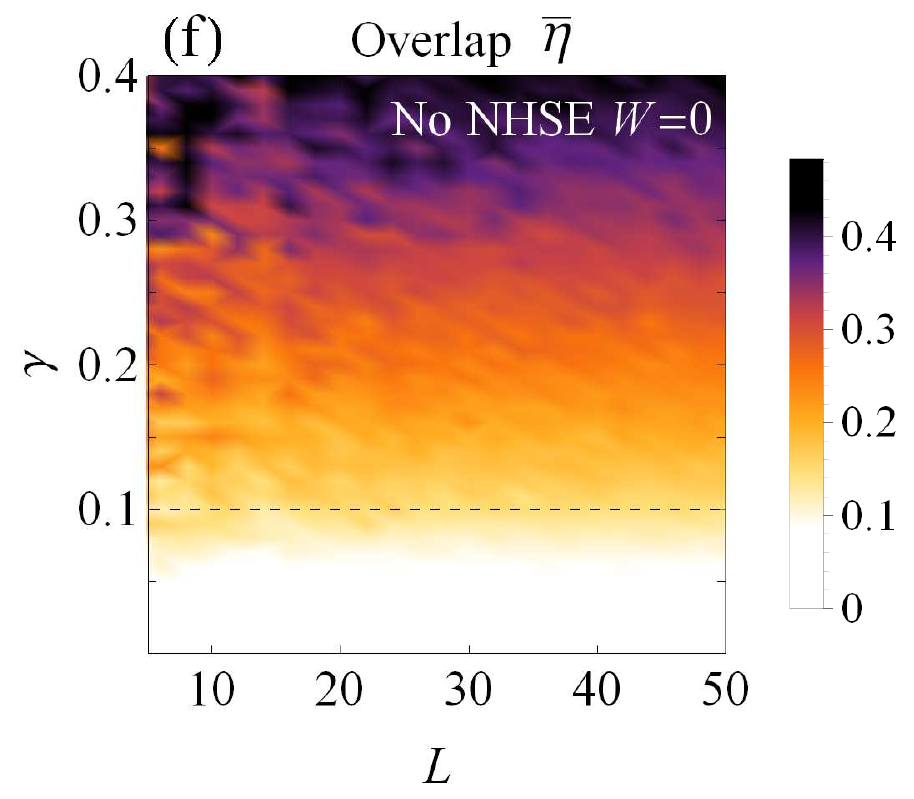}
\caption{Complex eigenenergies proportion $P$ and wavefunction overlap $\bar{\eta}$ for $L\times L$ squares. (a) $P$ for the single-band model [Eq. (\ref{H})], with $t=1$ and $s=0.3$. (b) The same as (a) except that onsite disorder is added with $W=0.05$, and the data is the average from ten disorder configurations.  (c) $P$ for the NHSE-free model [Eq. (\ref{H2})], with $m=0.5, t=0.2, \Delta=0$. Numerically, eigenenergies with imaginary part  $|\text{Im}(E)|>10^{-10}$ are regarded as complex. (d),(e),(f) The wavefunction overlap $\bar{\eta}$ corresponding to (a),(b),(c), respectively. The dashed line represents $\gamma L=1$ in (b)(e) and $\gamma=0.1$ in (c)(f).  } \label{singleband}
\end{figure}

To have a more complete picture, we calculate the complex proportion $P=N_c/N$ with varying $\gamma$ and $L$, where $N_c$ and $N$ denotes the number of complex-valued eigenenergies and all eigenenergies, respectively [Fig.~\ref{singleband}(a)]. While PT transition generally occurs as $\gamma$ increases for any fixed $L$, it occurs at smaller $\gamma$ for larger $L$. To see the robustness of this trend, we add weak random onsite disorder to the Hamiltonian \bea H'=H+\sum_{x,y} w(x,y)|x,y\ra\la x,y|  \label{disorder} \eea with $w(x,y)$ uniformly distributed in $[-W/2,W/2]$. Fig.~\ref{singleband}(b) shows that the trend is enhanced by disorder.

To demonstrate the role of NHSE, we consider a NHSE-free model as a comparison:
\bea
H(\bk)=(m+t\cos k_x+t\cos k_y)\sigma_z+\Delta\sigma_y+i\gamma\sigma_x, \label{H2}
\eea
where $\sigma_{x,y,z}$ are the Pauli matrices and all parameters are real-valued. The eigenvalues are $E_\pm(\bk)=\pm\sqrt{(m+t\cos k_x+t\cos k_y)^2+\Delta^2-\gamma^2}$. This Bloch Hamiltonian also has a generalized PT-symmetry $AH(\bk)A=H(\bk)$ where $A=\sigma_z\mathcal{K}$ and $A^2=1$\cite{Bender2002}. Apparently, PT symmetry is unbroken when $\gamma<\text{min}[\sqrt{(m+t\cos k_x+t\cos k_y)^2+\Delta^2}]$. Without NHSE, the OBC energies are similar to the PBC (Bloch) energies, and exhibit a nonzero PT threshold that is almost independent of $L$ [Fig.~\ref{singleband}(c)].

\emph{Non-perturbative mechanism and non-Hermiticity-size product.--}To understand the mechanism underlying Fig.~\ref{fig1}, we consider the overlap between a pair of eigenstates:
\bea
\eta(n,m)=\frac{|\la \psi_{n}|\psi_{m}\ra|}{\sqrt{\la \psi_{n}|\psi_{n}\ra\la \psi_{m}|\psi_{m}\ra}}, \label{eta}
\eea
where $H|\psi_{n}\ra=E_n|\psi_{n}\ra$, and similar for $m$.  We order all the real eigenvalues as $E_1\leq E_2\leq \cdots\leq E_{N_R}$. When an adjacent pair $(i,i+1)$ undergo the real-to-complex transition, the two eigenvectors become parallel to each other and we have $\eta(i,i+1)=1$ \cite{Ozdemir2019review}. Therefore, a measure of the transition tendency is the mean value of overlaps
\bea
\bar{\eta}=\frac{1}{N_R-1}\sum_{i=1}^{N_R-1} \eta(i,i+1). \label{mean}
\eea When $\bar{\eta}$ is closer to $1$, the real-to-complex transitions tend to occur more frequently as $\gamma$ increases.
The numerical $\bar{\eta}$ for model Eq.~(\ref{H}) is shown in Fig. \ref{singleband}(d,e), which exhibits very similar trend as the corresponding complex-value proportion $P$ in Fig.~\ref{singleband}(a,b). Thus, $\eta(n,m)$ and $\bar{\eta}$ contain useful information of the transition.

Now we exploit this information to understand the intriguing size-dependent behaviors. It should be explained why a very small $\gamma$ can induce $\eta(n,m)=1$ for many of the $(n,m)$'s, given that the reference point $\gamma=0$ guarantees $\eta(n,m)=0$ ($n\neq m$). We express the real-space Hamiltonian as $H=H_0+i\gamma V$, where $H_0$ and $V$ are both Hermitian. We treat the non-Hermitian term $i\gamma V$ as a perturbation, so that the eigenstates read
\bea
|\psi_{n}\ra=|\psi_n^{(0)}\ra+i\gamma\sum_{l\neq n}\frac{|\psi_l^{(0)}\ra\la \psi_l^{(0)}|V|\psi_n^{(0)}\ra}{E_n^{(0)}-E_l^{(0)}}+O(\gamma^2),
\eea
where $\{|\psi_n^{(0)}\ra\}$ are the unperturbed eigenstates spanning an orthonormal basis. By definition in Eq.~(\ref{eta}), we have
\bea
\eta(n,m)\approx 2\gamma  |\frac{\la \psi_n^{(0)}|V|\psi_m^{(0)}\ra}{E_n^{0}-E_m^{0}}|  \approx 2|\la \psi_n^{(0)}|\psi_{m}\ra|. \label{nm}
\eea For the simpler case of PBC, $n$ and $m$ has definite wavevector $\bk,\bk'$, respectively, and $\eta(n,m)$ is proportional to $|\la u_n^{(0)}(\bk)|u_{m}(\bk')\ra|\delta_{\bk,\bk'}$, where $|u\ra$ stands for the Bloch wavefunction. Thus, $\eta(n,m)=0$ when $\bk\neq\bk'$, and $\eta(n,m)\sim\gamma$ when $\bk=\bk'$. This perturbation picture breaks down only at Bloch-band-gap closing, when $E_n^{(0)}(\bk)-E_m^{(0)}(\bk)=0$ and $\eta(n,m)$ could be large according to Eq.~(\ref{nm}). Requiring multiple bands, this conventional scenario is irrelevant to our single-band model. Consequently, a small $\gamma$ generally cannot drive $\eta(n,m)$ to $1$ and cause PT breaking. For an OBC system without NHSE, the energy spectrum is asymptotically the same as in the PBC case for large size, so the same conclusion holds true.

Crucially, this perturbation approach breaks down for OBC systems with NHSE. In this case, the eigenstate $|\psi_{m}\ra \sim\exp(\kappa_x x+\kappa_y y)|\psi_m^{(0)}\ra\approx (1+\kappa_x x+\kappa_y y)) |\psi_m^{(0)}\ra$, where $\kappa_{x,y}$ is of order $\gamma$ (normalization of $|\psi_{m}\ra$ is unimportant because it only has higher-order contributions to the following results). Therefore, it follows from Eq.~(\ref{nm}) that $\eta(n,m)\approx 2|\la \psi_n^{(0)}|(\kappa_x x+\kappa_y y)|\psi_m^{(0)}\ra|$.    Since $x,y$ take values in $\{1,2,\cdots,L\}$, we expect that $\la \psi_n^{(0)}|x|\psi_m^{(0)}\ra$ and $\la\psi_n^{(0)}|y|\psi_m^{(0)}\ra$ can reach the order of $L$ for certain $(n,m)$. Hence, \bea \eta(n,m) \sim\gamma L. \label{L} \eea
The key feature is the presence of the $L$ factor, which would be absent without NHSE. Regardless of how small $\gamma$ is, it cannot be treated as a perturbation as size increases to $L\sim 1/\gamma$, otherwise we would have the apparently wrong result $\eta(n,m)>1$. While Eq.~(\ref{L}) is not supposed to be quantitatively precise, it offers a qualitative understanding. It means $\eta(n,m)\sim 1$ when $\gamma\sim 1/L$, and therefore $\eta(n,m)=1$ is possible, enabling the PT breaking. Numerically, we indeed see that $\gamma\sim 1/L$ is a visible characteristic scale of PT breaking [see the dashed line in Fig.~\ref{singleband}(b)(e)]. Intuitively, Eq.~(\ref{L}) suggests the non-Hermiticity-size product $\gamma L$ as a measure of the effective non-Hermiticity strength. Thus, the weakness of non-Hermiticity requires $\gamma L$ being small, in addition to the usual requirement of $\gamma/t$ being small.

Now we turn to the NHSE-free model Eq.~(\ref{H2}), for which the Bloch band theory applies and the OBC energies are asymptotically the same as the PBC ones. For PBC, $E_i$ and $E_{i+1}$ generally correspond to different $\bk$, so that the overlap $\eta(i,i+1)$ vanishes. Thus, Eq.~(\ref{mean}) is not a good measure of PT breaking tendency in the absence of NHSE. Instead, we define $\bar{\eta}=\sum_{i=1}^{N_{R+}} \eta(i,i')/N_{R+}$ for the model Eq.~(\ref{H2}), where $E_{i'}=-E_i<0$ and $N_{R+}$ is the number of positive real eigenenergies. For PBC, the opposite eigenenergies $E_+(\bk)$ and $E_-(\bk)=-E_+(\bk)$ share the same $\bk$, so that their wavefunction overlap can be nonzero as non-Hermiticity is turned on. Since the Bloch band theory applies, this definition is expected to remain informative under OBC. One can see this from Fig. \ref{singleband}(f), in which the trend of $\bar{\eta}$ is similar to that of $P$ in Fig.~\ref{singleband}(c).

We emphasize that for Bloch bands, modes with different $\bk$ cannot be coupled by non-Hermitian perturbations without breaking the translational symmetry. Hence, PT breaking can only occur in multi-band systems where non-Hermitian terms can couple different bands with the same $\bk$. Without NHSE, taking OBC would not alter this conclusion as the PBC and OBC energy bands are identical. In contrast, the non-Bloch PT breaking can occur even for single-band models, which indicates its different origin. For non-Bloch bands, energetically adjacent eigenstates from the same band are driven by the NHSE to be more parallel to each other as size increases, which enables the PT symmetry breaking. Since the entire band is involved, $P$ can be of order unity even for a small $\gamma\sim 1/L$. This sharply differs from the Bloch PT breaking where, irrespective of the size, $P$ is at most proportional to $\gamma$ \cite{Barashenkov2013,Konotop2016nonlinear}.

\begin{figure}
\includegraphics[width=4cm, height=3.45cm]{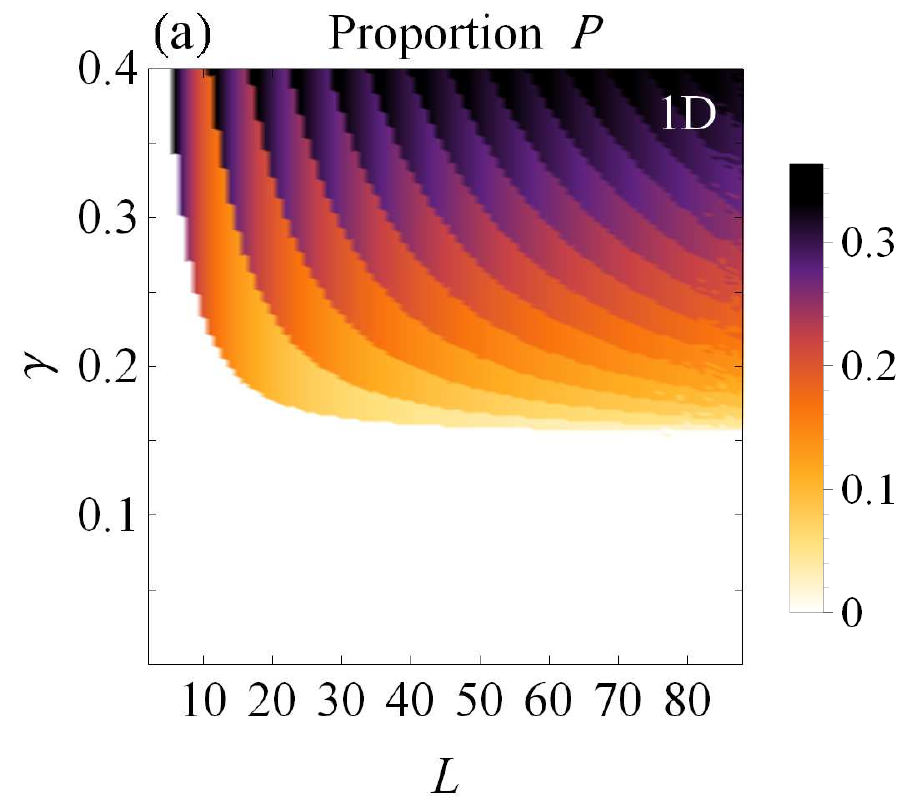}
\includegraphics[width=4cm, height=3.45cm]{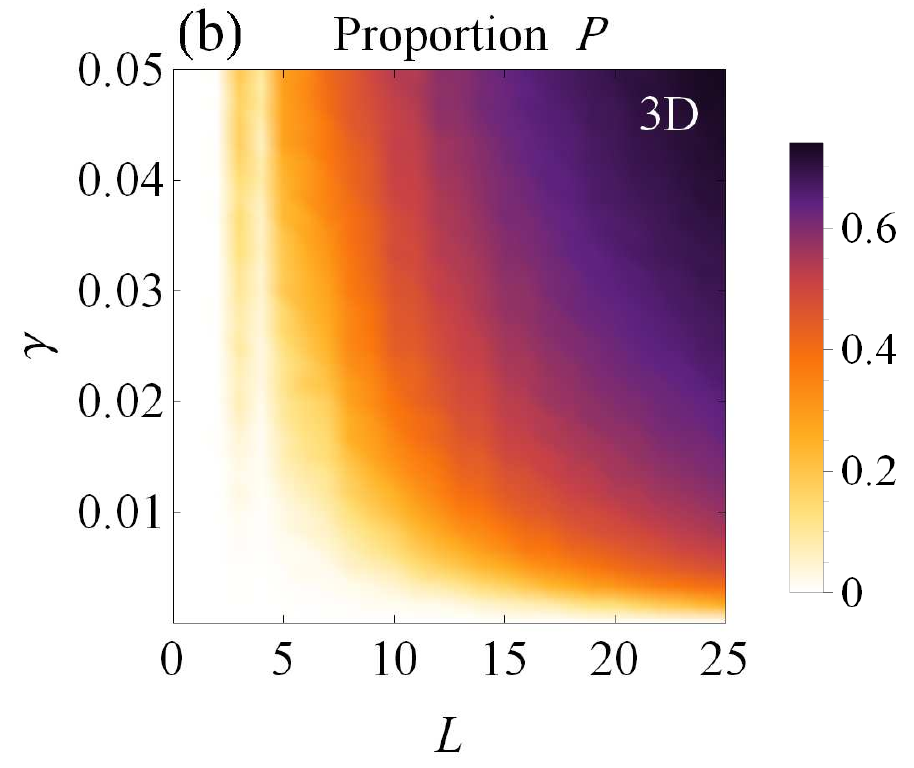}
\caption{Complex eigenenergies proportion $P$ for the 1D model and 3D cubic model (see text). (a) $P$ for $H_\text{1D}$ on open chains with length $L$. $t=1$, $s=0.15$.  (b) $P$ for $H_\text{3D}$ on $L\times L\times L$ cubes. $t=1$, $s=0.5$. For (b), onsite disorder is added akin to Eq. (\ref{disorder}), but only on boundary sites with $W=0.7$. The data is the average from six disorder configurations.    } \label{dimensions}
\end{figure}

\emph{Dimensional surprise.--}Perhaps the most unexpected aspect of our finding is the dependence on spatial dimensions. In the mechanism outlined above, NHSE appears to be the only crux of the matter, irrespective of the dimension. However, it is known for 1D non-Bloch PT breaking that the threshold does not approach zero as size increases \cite{Longhi2019nonBloch,Longhi2019Probing,Xiao2020nonBloch}. For example, we consider a 1D Hamiltonian $H_\text{1D}(k)=(t-\gamma)e^{ik}+(t+\gamma)e^{ik}+2s \cos 2k$ under OBC.  As shown in Fig.~\ref{dimensions}(a), the threshold approaches a nonzero constant as the size increases. Similar behaviors are known for the non-Hermitian SSH model \cite{yao2018edge,yin2018ssh}.

The puzzling difference between 1D and 2D can be understood as follows.  Eq.~(\ref{L}) means that $\eta(n,m)\sim 1$ at a small $\gamma\sim 1/L$, but it does not enforce $\eta(n,m)=1$. It is possible that the largest $\eta(n,m)$ approaches $1$ as size increases, but never equal to $1$. It turns out that it is indeed the case for 1D non-Bloch PT symmetry. The stability of real eigenenergies as size increases is closely related to a recently proved theorem stating that when $L\rw\infty$ the OBC eigenenergies form lines or arcs enclosing no area in the complex energy plane, while the Bloch $H(k)$ eigenenergies form loops \cite{Zhang2020correspondence,Okuma2020}. A line can be entirely on the real axis, meaning that PT symmetry can be exact at large size. Since this theorem stems from the 1D GBZ properties that are not generalizable to higher dimensions \cite{Zhang2020correspondence}, we believe that the 2D physics here is more generic and intrinsic to non-Bloch PT symmetry.

To confirm this understanding, we study a 3D model $H_\text{3D}(\bk)=\sum_{i=x,y,z}[(t-\gamma)e^{ik_i}+(t+\gamma)e^{-ik_i}]+8s\cos k_x\cos k_y \cos k_z$. From Fig.~\ref{dimensions}(b), we see that the threshold decreases towards $0$ as size increases, which is similar to 2D rather than 1D. Thus, the non-Bloch PT symmetry in spatial dimensions higher than one shares similar universal features. Note that we exclude fine-tuned cases that can be similarity-transformed to a Hermitian model, such as the $s=0$ case of Eq.~(\ref{H}). At the fine-tuned points, the behavior of $\eta(n,m)$ resembles that of 1D.

\begin{figure}
\includegraphics[width=4cm, height=3.45cm]{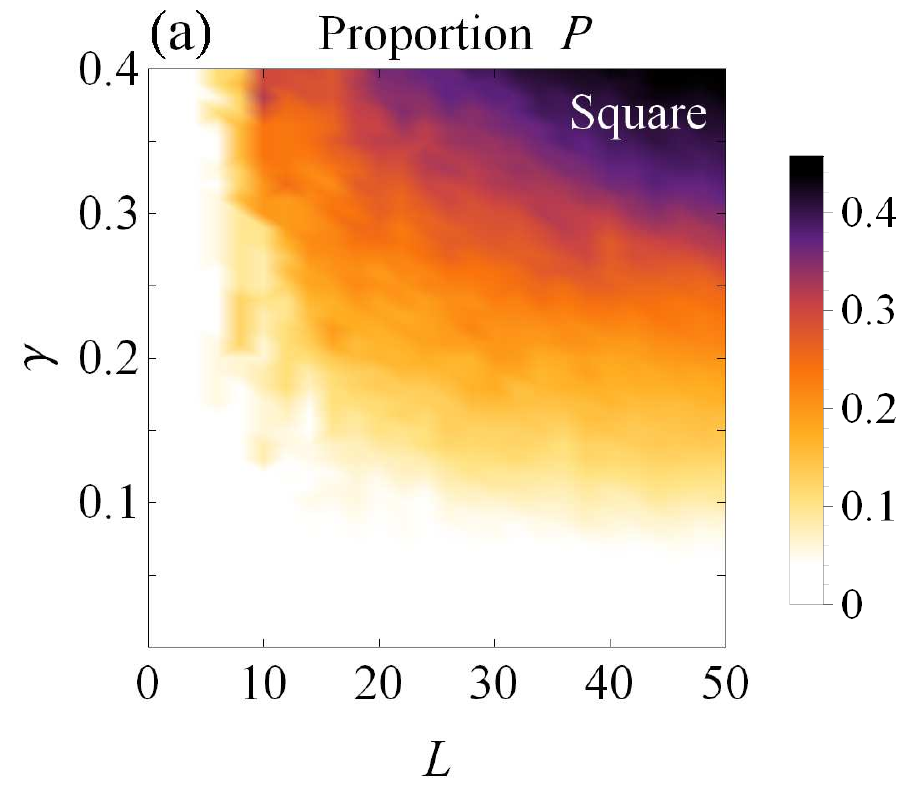}
\includegraphics[width=4cm, height=3.45cm]{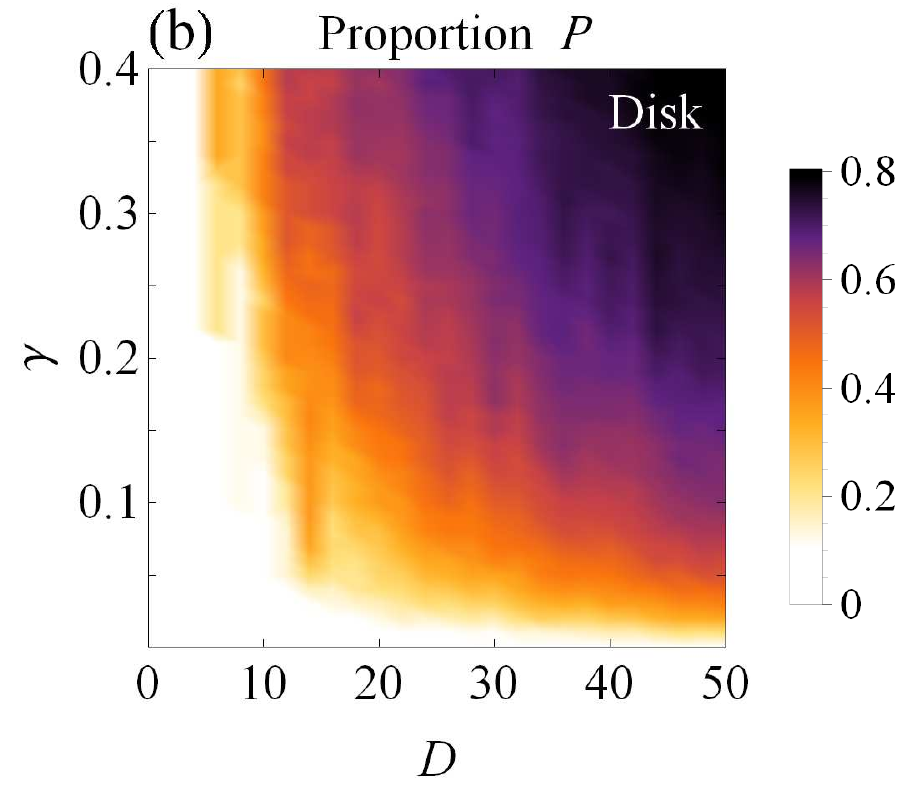}
\caption{The complex eigenenergies proportion of non-Hermitian Chern bands. $m=1.4$. (a) Square geometry with side length $L$. (b) Disk geometry with diameter $D$. } \label{Chernmodel}
\end{figure}

\emph{Non-Hermitian Chern bands.--}To further demonstrate the generality of the phenomenon, we consider a non-Hermitian Chern model in 2D \cite{yao2018chern,shen2017topological}:
\bea
H(\bk)=&&(\sin k_x+i\gamma)\sigma_z+\sin k_y\sigma_y \nn\\ && +
(m+\cos k_x+\cos k_y)\sigma_x.  \label{Chern}
\eea
The non-Bloch PT symmetry at small size has been noticed before \cite{yao2018chern}, but its breaking at larger size was not touched. Here, we calculate the size dependence of $P$ for the square geometry, and find that $P$ increases as size increases [Fig.~\ref{Chernmodel}(a)].  This trend is even stronger for the disk geometry [Fig.~\ref{Chernmodel}(b)], though different geometries share qualitatively similar behavior. This is reasonable as the square geometry is more ``regular'' than the disk, leading to the suppression of certain matrix elements
$\la\psi_n^{(0)}|V|\psi_m^{(0)}\ra$ [see Eq.~(\ref{nm})]. In fact, after adding weak disorder as in Eq.~(\ref{disorder}) to the square to break the symmetry, we find that $P$ is significantly increased, resembling that of the disk geometry.

Finally, we emphasize that Eq.~(\ref{Chern}) only involves onsite gain/loss, meaning that nonreciprocal hopping is not a necessary ingredient to  induce the described phenomenon.

\emph{Conclusions.--}We have uncovered generic behaviors of non-Bloch PT symmetry in spatial dimensions higher than one. In the presence of NHSE, we find that the product of bare non-Hermiticity and system size is a measure of the effective non-Hermiticity, meaning that even a weak non-Hermiticity becomes effectively strong as size increases. This non-perturbative effect causes the asymptotic vanishing of PT threshold, which is a universal property of non-Bloch PT breaking in two and higher dimensions. Notably, 1D systems evade the above physics, suggesting rich and unexpected interplay between non-Bloch physics and spatial dimensionality. Our theory is testable on several experimental platforms. For example, the nonreciprocal model Eq. (\ref{H}) can be realized in topolectrical circuits \cite{Helbig2019NHSE}, and the onsite-non-Hermiticity model Eq. (\ref{Chern}) can be constructed in coupled ring resonators with gain/loss \cite{Hafezi2011robust,bandres2018topological}.

{\it Acknowledgements.--} This work is supported by NSFC under Grant No. 11674189.

\bibliography{dirac}

\end{document}